\documentclass[fleqn,usenatbib]{mnras}

\usepackage{newtxtext,newtxmath}

\usepackage[T1]{fontenc}
\usepackage{ae,aecompl}

\usepackage{graphicx}	
\usepackage{amsmath}	
\usepackage{amssymb}	
\usepackage[figure,figure*]{hypcap}
\usepackage{array,multirow,graphicx}

\newcommand{\fesc}{f_{\mathrm{esc}}}
\newcommand{\xiion}{\xi_{\mathrm{ion}}}

\newcommand{\EWHB}{\mathrm{EW}(\Hbeta)}
\newcommand{\Hbeta}{\mathrm{H} \beta}

\newcommand{\Otwo}{\mathrm{O \footnotesize{II}} }
\newcommand{\Othree}{\mathrm{O  \footnotesize{III}}}

\title[LyC leakage versus quenching with the JWST]{Lyman continuum leakage versus quenching with the James Webb Space Telescope: The spectral signatures of quenched star formation activity in reionization-epoch galaxies}

\author[C. Binggeli at al.]{Christian Binggeli$^{1}$\thanks{Christian.binggeli@physics.uu.se},
Erik Zackrisson$^{1}$,
Kristiaan Pelckmans$^{2}$,
Rub\'{e}n Cubo$^{2}$,
\newauthor
Hannes Jensen$^{1}$,
Ikko Shimizu$^{3}$
\\
$^{1}$Department of Physics and Astronomy, Uppsala University,
Box 515, SE-751 20 Uppsala, Sweden\\
$^{2}$Department of Information Technology, Division of Systems and Control,
Uppsala University, Box 337, SE-751 05 Uppsala, Sweden\\
$^{3}$Theoretical Astrophysics, Department of Earth and Space Science, Osaka University,
1-1 Machikaneyama, Toyonaka, Osaka 560-0043, Japan
}

\date{Accepted 2018 April 20. Received 2018 April 04; in original form 2018 January 24}

\pubyear{2018}

\begin{document}
\label{firstpage}
\pagerange{\pageref{firstpage}--\pageref{lastpage}}
\maketitle

\begin{abstract}
In this paper, we study the effects of a recent drop in star formation rate (SFR) on the spectra of epoch of reionization (EoR) galaxies, and the resulting degeneracy with the spectral features produced by extreme Lyman continuum leakage. 
In order to study these effects in the wavelength range relevant for the upcoming James Webb Space Telescope (JWST), we utilize synthetic spectra of simulated EoR galaxies from cosmological simulations together with synthetic spectra of partially quenched mock galaxies. We find that rapid declines in the SFR of EoR galaxies could seriously affect the applicability of methods that utilize the equivalent width of Balmer lines and the ultraviolet spectral slope to assess the escape fraction of EoR galaxies. In order to determine if the aforementioned degeneracy can be avoided by using the overall shape of the spectrum, we generate mock NIRCam observations and utilize a classification algorithm to identify galaxies that have undergone quenching. We find that while there are problematic cases, JWST/NIRCam or NIRSpec should be able to reliably identify galaxies with redshifts $z\sim 7$ that have experienced a significant decrease in the SFR (by a factor 10-100) in the past 50-100 Myr with a success rate $\gtrsim 85\%$. We also find that uncertainties in the dust-reddening effects on EoR galaxies significantly affect the performance of the results of the classification algorithm. We argue that studies that aim to characterize the dust extinction law most representative in the EoR would be extremely useful.
\end{abstract}




\section{Introduction}
\label{section:introduction}
The epoch of reionization (EoR) represents an extremely important yet poorly understood stage in the evolution of the Universe. During the EoR, the Universe was flooded by energetic hydrogen-ionizing Lyman continuum radiation (LyC) which ionized the neutral intergalactic medium (IGM), but the source of this radiation remains to be determined. Star-forming galaxies have emerged as the main candidate for producing the bulk of the ionizing photons required to sustain reionization, but it is not clear whether the early star-forming galaxies produced and leaked enough such radiation.

In the low- and intermediate-redshift Universe, studying the leakage of LyC photons can be done directly by detecting the escaping LyC photons. There are now several studies that detect leaking LyC with $\fesc \sim 0.01-0.1$ \citep[e.g.][]{bergvall_first_2006,leitet_escape_2013, borthakur_local_2014,izotov_eight_2016,izotov_detection_2016} at redshifts $z\lesssim 2$.  There are also studies that claim extreme ($\fesc > 0.4$) leakage \citep{vanzella_hubble_2016,shapley_q1549-c25:_2016,bian_high_2017, izotov_J1154_2018}. \citet{matthee_production_2017} also present very high escape fractions ($\fesc \approx 0.3-0.45$) for a number of individual sources in a large sample of Lyman-$\alpha$ and H$\alpha$ emitters at $z\sim 2$. When stacking the whole sample they find escape fractions $f_\mathrm{esc}<0.1$ for both median and mean stacking. However, detecting the LyC photons directly becomes impossible at $z\gtrsim 4-5$ due to absorption in the increasingly neutral IGM as we approach the EoR. A possible workaround for this is to indirectly determine the LyC escape fraction ($\fesc$) using parts of the spectrum that are unaffected by IGM absorption. A couple of methods for indirectly determining the escape fraction have been suggested. 
A method using the strengths of absorption lines to probe the line-of-sight covering fraction of neutral hydrogen was proposed by \citet{jones_keck_2013} and \citet{leethochawalit_absorption-line_2016}. The method has recently been successfully applied to known LyC leakers by \cite{gazagnes_neutral_2018} and \cite{chisholm_accurately_2018}. This method is, however, limited by signal-to-noise requirements at sufficient spectral resolution.
It has been suggested that high $\left[ \Othree \right]/\left[ \Otwo \right]$ line ratios could be indicative of Lyman continuum leakage through density-bounded regions in local compact star-forming galaxies \citep{jaskot_origin_2013}. Low-redshift star-forming galaxies selected for high ratios and compactness have been found to show significant escape fractions \citep{izotov_eight_2016,izotov_detection_2016,izotov_J1154_2018}, see also \citet{chisholm_accurately_2018}. Meanwhile, \citet{stasinska_excitation_2015} have shown that high $\left[ \Othree \right]/\left[ \Otwo \right]$ ratios are not necessarily an indicator of leaking LyC.

\citet{verhamme_using_2015} propose that the shape of the Lyman-$\alpha$ spectral line may hold information about the escape fraction, which observations of Lyman-$\alpha$ profiles of local LyC emitters seem to support \citep{verhamme_using_2015,verhamme_lyman_2017}. However, as the neutral fraction of the IGM increases with redshift, the shape of the Lyman-$\alpha$ line may change due to Lyman-$\alpha$ transmission effects. Thus, it is uncertain how applicable this method will be to reionization-epoch galaxies \citep{verhamme_lyman_2017}.

Another method was suggested by \citet{zackrisson_spectral_2013}, and later tested on simulated galaxies from cosmological simulations in \citet{zackrisson_spectral_2017}. This method exploits the fact that ionizing radiation produced by a young hot population of stars will become reprocessed to longer wavelengths in the neutral hydrogen gas surrounding newly formed stars. Thus, the LyC photons will leave an imprint in the rest-frame non-ionizing UV/optical parts of the spectrum. The authors argue that this could in the future be observed using the James Webb Space Telescope (JWST). By measuring the equivalent width of the Balmer beta line ($\Hbeta$) while simultaneously measuring the ultraviolet spectral slope, one should be able to identify cases of extreme LyC leakage ($\fesc \gtrsim 0.5$) from reionization-epoch galaxies \citep{zackrisson_spectral_2017}. In \citet{jensen_machine-learning_2016}, this method was further expanded upon using machine learning algorithms in order to increase the number of spectral features used to predict the escape fraction using mock JWST/NIRSpec observations. For a discussion on the caveats of this method, see \citet{zackrisson_spectral_2017}

Since there are indications that star formation rates (SFR) in galaxies in the EoR are generally increasing over time \citep[e.g.][]{finlator_smoothly_2011}, optical recombination lines of these galaxies are expected to be strong. The method by \citet{zackrisson_spectral_2013} utilizes this, and therefore, any weak emission lines in galaxies with blue UV slopes will be interpreted as a telltale sign of Lyman continuum leakage. This could, however, make the method sensitive to rapid changes in star formation. Since the emission in these lines depends on the reprocessing of LyC, and LyC production is dominated by the short-lived O and B stars, a drop in the SFR can lead to weaker lines while not necessarily leading to a significantly redder UV slope. While many cosmological simulations predict star formation histories with fairly modest fluctuations in SFR in the early Universe \citep{finlator_host_2013, shimizu_physical_2014,gnedin_cosmic_2014-1,gnedin_cosmic_2014}, there are simulations that do feature such variations even for galaxies with stellar masses above $\mathrm{M_{\star}}=10^8 \; \mathrm{M_{\sun}}$ \citep{kimm_towards_2015,ma_difficulty_2015,ma_simulating_2017}. Furthermore, the recent discovery of a $z \approx 9.1$ galaxy that may have undergone a significant drop in the SFR \citep{hashimoto_onset_2018} raises the question of how common these types of galaxies may be in the early Universe.

If one would catch a galaxy soon after it has undergone a rapid drop in SFR, this object could be incorrectly identified as a galaxy with high LyC leakage by the method described in \citet{zackrisson_spectral_2013,zackrisson_spectral_2017,jensen_machine-learning_2016}. 

We utilize simulated galaxies from the \citet{shimizu_physical_2014} cosmological simulation in combination with mock-galaxies that have experienced a recent drop in star formation activity in order to study the effect of quenching on the applicability of using emission lines and the UV slope as a probe of the LyC escape fraction. We assess the possibility of identifying quenched galaxies during the EoR using JWST/NIRCam and MIRI photometry. Finally, we utilize linear discriminant analysis (LDA)\citep[][pp.106-112]{hastie_elements_2009} as a classification algorithm in order to assess how well we will be able to distinguish between galaxies that have experienced a recent decline in SFR and star-forming galaxies with high escape fractions. We also asses in which cases the information in the spectral energy distribution is insufficient to realistically distinguish the two types.
Note that we do not strictly distinguish between `attenuation' and `extinction', but use the terms interchangeably when talking about dust-reddening effects.

\section{Model}
Here, we present the models used in this paper and some important properties of these models. More in-depth properties of the model are discussed in \citet{zackrisson_spectral_2017}. 

\begin{figure}
	\centering 
        \includegraphics[width=0.52\textwidth]{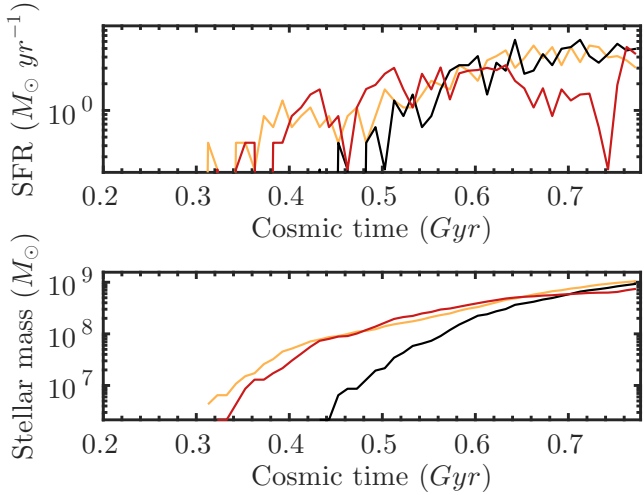}

    \caption{\emph{Top:} Star formation histories binned in 10 Myr bins over cosmic time ($t=0$ to the age of the Universe at $z=7$) for two typical (black and yellow) and one extremely stochastically star-forming galaxy \emph{(red)} from the \citet{shimizu_physical_2014} simulations. \emph{Bottom:} Stellar masses for the same three galaxies over cosmic time.}
    \label{figure:SFH_Shimizu}
\end{figure}

\subsection{Simulated galaxies and SED modeling}
\label{sec:simulations}
We use the cosmological hydrodynamic simulation by \citet{shimizu_physical_2014}, which constitutes a subset of the galaxies used in \citet{zackrisson_spectral_2017}. The simulation is based on the smoothed particle hydrodynamic code {\scriptsize GADGET-3} which is an improved version of the public {\scriptsize GADGET-2} code \citep{Springel_cosmological_2005}. In the simulation, $2 \times 640^3$ dark matter and gas particles are simulated in a comoving volume of $50 h^{-1}$ Mpc cube while considering star formation, supernova feedback and chemical enrichment following \citet{Okamoto2008}. The simulation does not resolve single stars, but rather collections of stars with masses $\approx 10^6 M_{\sun}$ with a Salpeter IMF \citep{salpeter_luminosity_1955}, which leads to a minimum stellar mass for the simulated galaxies of around $10^7 M_{\sun}$.
We extract the star formation histories, internal metallicity distribution, and the dust content of 406 galaxies with stellar masses $M_{\star} \geq 5\times 10^{8}\; M_{\sun}$ at redshift $\mathrm{z} \approx 7$ and dust-free magnitudes in the range $M_{1500,AB} \approx -24$ to $-20$ (corresponding to apparent magnitudes $\mathrm{m_{1500,AB}} \approx 23-27$ at $z=7$) from the simulation.

The information about the star formation history and metallicity stored in star particles representing a collection of stars with mass of $M_{\star}\approx 10^{6}\; M_{\sun}$, while the dust content is calculated for each galaxy as a whole in the simulation and given as a prediction of the extinction at 1500~\AA~($\mathrm{A_{UV}}$).

The average stellar metallicity of the galaxies in the simulation ranges from $\mathrm{Z = 7 \times 10^{-4}}$ to  $\mathrm{Z = 6 \times 10^{-3}}$, while single SPH particles can have metallicities $\mathrm{Z<10^{-5}}$. In \autoref{figure:SFH_Shimizu}, we show the star formation histories and total stellar mass as a function of cosmic time for three galaxies from the simulations. On average, the galaxies show increasing SFRs with small variations (factor 2-3) on 10 Myr time-scales once the galaxies reach stellar masses around $\mathrm{10^{8}\; M_{\sun}}$, with $\sim 5\%$ of galaxies displaying larger variations (factor 5-10). This is shown in \autoref{figure:SFH_Shimizu}, where the black and yellow lines show two typical galaxies, while the red line shows the most extreme galaxy extracted at $\mathrm{z=7}$. This galaxy has a rapid change in SFR (about a factor 10 drop and shortly thereafter a factor 20 increase) at around 0.74 Gyr after Big Bang. This type of galaxy represents an extremely rare type of galaxy in the \citet{shimizu_physical_2014} simulation. However, variations of this size and even more extreme cases, with fluctuations in SFR of the order of $\sim 10-100$ over $\sim 10-100$ Myr, are seen more routinely in other simulations \citep{kimm_escape_2014,kimm_towards_2015,ma_difficulty_2015,ma_simulating_2017,trebitsch_fluctuating_2017}. Variations of this kind would also be a consequence of a star formation duty cycle of  $\sim 10$ \% as advocated by \citet{wyithe_predicted_2014}.

In this paper, the term `\emph{quenching}' is used to describe sudden a drop in the SFR, in contrast to the more regular usage for the term describing a total shut-down in star formation in a galaxy \citep{harker_population_2006,faber_galaxy_2007} . The drop in star formation can be explained for example by the removing of gas due to feedback effects after a recent starburst event \citep{ma_simulating_2017}

In order to generate quenched galaxies, we extract simulated galaxies from the \citet{shimizu_physical_2014} set and lower the star formation with a fixed factor 5,10 and 100 and simulate continuous star formation for 10-100 Myr by aging the already existing population while adding a less massive population of stars with ages distributed evenly over the time after the quenching takes place and metallicities equal to the average of the simulated stellar population. Thus, the quenched galaxies will have a range of internal metallicity distributions and varying star formation histories up until a significant drop in the SFR occurs. 
The values for the drop in the SFR and the duration of the quenching were selected to be roughly of the same size as those observed in simulations by \citet{kimm_towards_2015} and \citet{ma_difficulty_2015,ma_simulating_2017}, and should be comparable to to the kind of SFR fluctuations seen in simulations by \citet{kimm_escape_2014,trebitsch_fluctuating_2017}.

The resulting galaxies are fainter than the simulated ones with dust-free magnitudes in the range $M_{1500,AB}\approx -23.5$ to $-17.5$. In the simulation by \cite{shimizu_physical_2014}, this kind of galaxy that is caught in the low SFR stage after quenching is rarer than one in a hundred. In principle, we expect that the effect of this type of quenching on the SED will not be dissimilar to a less massive young starburst on top of a massive underlying aged population, as discussed in \citet{zackrisson_spectral_2013}. In that study, the authors argue that, for dust-free galaxies, continuum measurements with NIRSpec and MIRI could be used to identify such cases. This could however be complicated further when the difference in age between the massive and less massive population is smaller or when taking more complex star formation histories and dust and into consideration. 

\begin{figure}
	
        \includegraphics[width=0.52\textwidth]{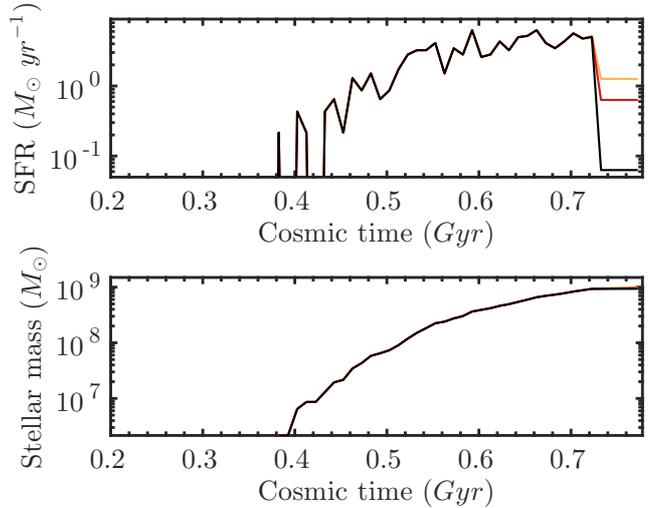}
    \caption{\emph{Top:} Star formation histories binned in 10 Myr size bins over cosmic time ($t=0$ to age of Universe at $z=7$) for a mock quenched galaxy shown with a factor of 5 \emph{(yellow)}, 10 \emph{(red)} and a factor of 100 \emph{(black)} drop in star formation over 50 Myr. \emph{Bottom:} Stellar masses for the same galaxies over cosmic time. Note that the way the mock galaxies are created they have a mismatch in age with the galaxies from the simulations, which grows with time after the quenching occurs.}
    \label{figure:SFH_Mock_Galaxies}
\end{figure}

In \autoref{figure:SFH_Mock_Galaxies}, we show the star formation history and total stellar mass assembly history for three quenched mock galaxies. We assume a constant SFR once the quenching has occurred.

To get spectral energy distributions (SEDs) from the collection of star particles and the dust content, we use a grid of synthetic spectra for metallicity ranging from $Z=10^{-7}$ to $Z=0.03$ generated with the Yggdrasil spectral evolutionary code \citep{zackrisson_spectral_2011}. This way, we can assign a spectrum to each star particle and sum over the star particles in the galaxies. To find the closest relevant spectrum for each particle, we perform an interpolation in $\mathrm{log(age)}$ and $\mathrm{log(Z)}$. The grid of spectra is generated using the BPASS v2.0 binary stellar evolutionary models \citep{eldridge_spectral_2009, stanway_stellar_2016} in combination with spectra from \citet{raiter_predicted_2010} for extremely metal poor stars ($Z=10^{-7}-10^{-5}$). 
We calculate the Lyman continuum photon production efficiency (the number of ionizing photons produced per unit UV luminosity, $\xiion$, in $\mathrm{erg}^{-1} \mathrm{Hz}$) for the simulated galaxies. We find values of $log_{10}(\xiion/ \mathrm{erg}^{-1} \mathrm{Hz}) \approx 25.2-25.6$ for the individual galaxies, with a mean value for the simulated galaxy set of $\langle \mathrm{log_{10}}( \xiion / \mathrm{erg^{-1}Hz})\rangle = 25.48 \;$. This result is consistent with several observational studies \citep[e.g.][]{stark_spectroscopic_2015,bouwens_reionization_2015,bouwens_lyman-continuum_2016}, motivating the use of the BPASS binary stellar evolutionary models \footnote{Note that the newer version of the BPASS binary evolution code does predict lower values of $\xiion$ \citep{eldridge_binary_2017}}.
The nebular emission associated with each star particle is calculated using the Cloudy photo-ionization code \citep{ferland_2013_2013} while taking the escape fraction into account and under the assumption that the nebular metallicity is equal to the stellar metallicity. See \citet{zackrisson_spectral_2013,zackrisson_spectral_2017} for more in-depth discussion of the procedure. To account for dust, we use the simulated dust amount from the \cite{shimizu_physical_2014} simulations and apply \cite{calzetti_dust_2000} reddening with the same attenuation for the nebular and stellar component ($\mathrm{E(B-V)_{stars}=E(B-V)_{neb}}$) as the fiducial model. With the amount of dust given in the simulations, this gives an average UV slope $\beta \approx -2.0$ for the whole sample, which is consistent with observations of $z\sim 7$ galaxies \citep{bouwens_UV_2014}. The mean optical extinction in the V-band for these galaxies is $\mathrm{ A_{V}} \approx 0.4 \ \mathrm{mag}$ with a few galaxies reaching as high as $\mathrm{ A_{V}} \approx 1 \ \mathrm{mag}$. In \autoref{sec:dust_handling} we discuss the effects of other attenuation and extinction laws, and we there use the standard \cite{calzetti_dust_2000} attenuation law (where $\mathrm{E(B-V)_{stars}=0.44E(B-V)_{neb}}$) and the \cite{pei_interstellar_1992} SMC extinction law. For the \cite{pei_interstellar_1992} SMC law, the mean optical extinction in the V-band is $\mathrm{ A_{V}} \approx 0.2 \ \mathrm{mag}$. While the Calzetti attenuation law gives UV slopes consistent with observations of $z\sim 7$ galaxies, and there are observations of Lyman break galaxies at $z\sim 5$ that point towards a Calzetti-like attenuation law \citep{koprowski_direct_2018}, we note that there are also recent observations that favor a steeper extinction curve for galaxies at $z\sim 7$ \citep{smit_measurement_2017}.

In this paper, we assume  that all leakage of LyC radiation is occurring through holes in the neutral gas envelope surrounding star-forming regions. This is what is described in \citet{zackrisson_spectral_2013} as the ionization-bounded nebula with holes (a.k.a. the picket-fence model). In this type of geometry, the Lyman continuum escape fraction is simply $\fesc=1-\mathrm{f_{cov}}$ where $\mathrm{f_{cov}}$ is covering fraction of neutral gas. This basically means that the nebular emission will be scaled according to the escape fraction. We calculate spectra for $\fesc = 0, 0.05, 0.10 \dotso 1.0$. This gives us a total of $\approx 250000$ quenched mock galaxies and a total of $\approx 8500$ simulated galaxies with varying escape fractions and different star formation histories.

\subsection{Mock observations}
\label{sec:mock_obs}
In order to test whether we can realistically distinguish between quenched and normally star-forming galaxies with variable escape fractions we create mock JWST/NIRCam and MIRI photometry observations of the galaxies and add observational noise to these. For the noise we assume Gaussian noise with a fixed signal-to-noise ratio of $\mathrm{S/N}=10$ in all filters independent of their sensitivity. 

In reality, the $\mathrm{S/N}$ will of course depend on the sensitivity in a given filter. For the NIRCam observations, we expect this level of $\mathrm{S/N}$ in all filters for around 30 minutes exposure time for a galaxy with $\mathrm{m}_{\mathrm{AB,1500}}=27$\footnote{Calculated using the JWST exposure time calculator (\url{https://jwst.etc.stsci.edu/})}. For MIRI, we expect a signal-to-noise ratio of at least 10 for galaxies with $\mathrm{m}_{\mathrm{AB,1500}}=27$ with an exposure of $~50$ hours in the F560W filter. Observations of this type will be performed within the JWST guaranteed time observations \footnote{For example, see GTO programs 1180 and 1283, \url{https://jwst-docs.stsci.edu/display/JSP/JWST+GTO+Observation+Specifications}}. In total, we use mock observations in six NIRCam filters (F115W, F150W, F200W, F277W, F356W and F444W) as well as the F560W MIRI filter.

\subsection{Linear discriminant analysis (LDA)}
\label{Sec:LDA}
The classification algorithm linear discriminant analysis (LDA) is used in order to classify the galaxies as quenched or as normally star-forming \citep[][pp.106-112]{hastie_elements_2009}. 

A classification algorithm as LDA aims to recover the probability of an object belonging to a class $C$ given input features $X$.
LDA is based on Bayes theorem, and makes the parametric assumption that the involved probabilities can be modeled as multivariate Gaussian distributions with common covariances but class specific means. 
Given those quantities, a sample with input features $x$ is assigned to class $k$ or $l$ according to the log-likelihood ratio:
\begin{equation}
	\label{eq:LDA_prob}	
\begin{aligned}
\mathrm{log}\frac{\mathrm{P}(G=k|X=x)}{\mathrm{P}(G=l|X=x)} ={} & \mathrm{log}\frac{\pi_k}{\pi_l}-\frac{1}{2}(\mu_k-\mu_l)^T\boldsymbol{\Sigma}^{-1}(\mu_k-\mu_l) \\
      & +x^T\boldsymbol{\Sigma}^{-1}(\mu_k-\mu_l)
\end{aligned}
\end{equation}
where $\mathrm{P}(G=k|X=x)$ and $\mathrm{P}(G=l|X=x)$ are the probabilities of an observation or object belonging to class $k$ and $l$ respectively, given input features $x$. $\pi$ are the prior probabilities of the classes, $\mu$ are the means and $\boldsymbol{\Sigma}$ denotes the common covariance matrix. \autoref{eq:LDA_prob} implies that the decision boundary between the classes (where the probabilities are equal) is linear in x. Classification to a class $k$ and $l$ is done according to the maximum conditional probability of belonging to a class. 

The training problem amounts to estimating the quantities $\mu_k,\mu_l$ and $\boldsymbol{\Sigma}$ that best describe the classes given some training set of input features. This is performed using maximum likelihood. In this work, the features are NIRCam and MIRI magnitudes and the classes are galaxies that have undergone quenching and those that have not. To avoid skewed data, we under-sample the quenched set of galaxies so that we end up with balanced sets of simulated star-forming galaxies and mock quenched galaxies.

\section{Quenched galaxies and observables}
\begin{figure}
\centering 
        \includegraphics[width=\columnwidth]{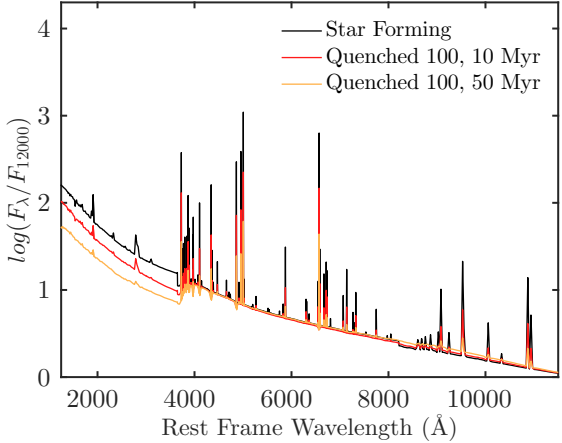}
    \caption{Synthetic spectrum of a single star-forming galaxy ($\mathrm{M_{\star}}\approx 7 \times 10^8 \; \mathrm{M_{\sun}}$) from the simulations \emph{(black)} and the same galaxy after quenching with a factor 100 has taken place for 10 \emph{(red)} and 50 \emph{(yellow)} Myr. The spectra have been normalized at 12000 \AA}
    \label{figure:SED_quenching}
\end{figure}

\autoref{figure:SED_quenching} shows the general effect that quenching has on the SED. As the number of Lyman continuum photons produced drops, the number of such photons that are reprocessed into optical and non-ionizing UV goes down. The emission lines become weaker and the continuum drops in the UV while remaining more or less unchanged at longer wavelengths. The net effect is a reddening of the continuum. 10 Myrs after quenching, $\EWHB$ has dropped a factor $\sim 4$ while the UV slope remains more or less the same as in the non-quenched case. 50 Myrs after quenching the UV slope has changed significantly while $\EWHB$ has dropped further. However, the UV slope does not experience extreme reddening due to the quenching. The average reddening of the UV slope over time is only $\Delta\beta \sim 0.2$ and $\sim 0.5$ for a factor of 10 and 100 drop in SFR over 100 Myr, respectively. While this slight change in the UV slope could be used to identify cases of quenching if dust effects are ignored, the reddening of the UV slope by dust if one assumes a Calzetti attenuation law for the galaxies is of the same order.The quenched galaxies would thus be indistinguishable from extreme leakers using only $\EWHB$ and the UV slope $\beta$ as diagnostics for escape fraction. 

\begin{figure*}
	\centering 
	\begin{tabular}{c}
        \includegraphics[width=\textwidth]{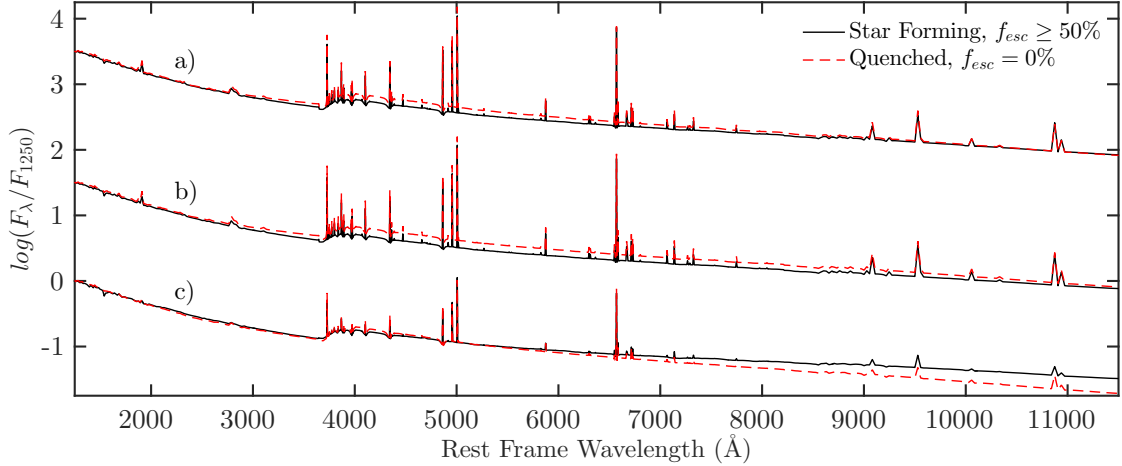}
	\end{tabular}
    \caption{Synthetic spectra showing three examples of the degeneracy between escape fraction and SFR in dusty star-forming galaxies (black) with extreme escape fractions and quenched galaxies (red) with zero escape fractions and lower dust content. \emph{a)}: Simulated star-forming galaxy with $\fesc=50\%$, $\mathrm{ A_{V} \approx 0.41 \; \mathrm{mag}}$ and mock galaxy quenched with a factor of 5 with $\mathrm{ A_{V} \approx 0.15 \; \mathrm{mag}}$. \emph{b)}: Simulated star-forming galaxy with $\fesc=70\%$, $\mathrm{ A_{V} \approx 0.49 \; \mathrm{mag}}$ and mock galaxy quenched with a factor 10 and $\mathrm{ A_{V} \approx 0.21 \; \mathrm{mag}}$. \emph{c)}: Simulated star-forming galaxy with $\fesc=90\%$, $\mathrm{ A_{V} \approx 0.51 \; \mathrm{mag}}$ and mock galaxy quenched with a factor 100 and $\mathrm{ A_{V} \approx 0.09 \; \mathrm{mag}}$. All quenched galaxies are observed 40 Myr after the quenching takes place and the Calzetti attenuation law, where $\mathrm{E(B-V)_{stars}=E(B-V)_{neb}}$ was used for all galaxies. For clarity, the spectra have been normalized at 1250 \ \AA, \emph{a)} and \emph{b)} have been shifted upwards.}
    \label{figure:SED_example_gal}
\end{figure*}

Examples of this degeneracy between extreme leakers and quenched galaxies are shown in \autoref{figure:SED_example_gal}. This figure displays three pairs of simulated and quenched galaxies, where the galaxies in each pair have basically identical $\EWHB$ and UV slopes. The pairs in \emph{a)} and \emph{b)} in this figure show excess flux in the continuum in JWST NIRSpec and NIRCam wavelengths. This is seen especially around 4000 \AA. In \emph{c)}, the galaxies look virtually identical in this wavelength range while the dustier star-forming galaxy is brighter in the rest-frame near-infrared relative to the quenched galaxy. In \autoref{sec:NIRCAM_MIRI_ML} we test whether or not these differences in the continuum are sufficient to identify quenched galaxies when observational noise is taken into account.

\subsection{NIRCam+MIRI to identify quenched galaxies}
\label{sec:NIRCAM_MIRI_ML}
\begin{figure*}
	\centering 
	\begin{tabular}{c c}
        \includegraphics[width=0.52\textwidth]{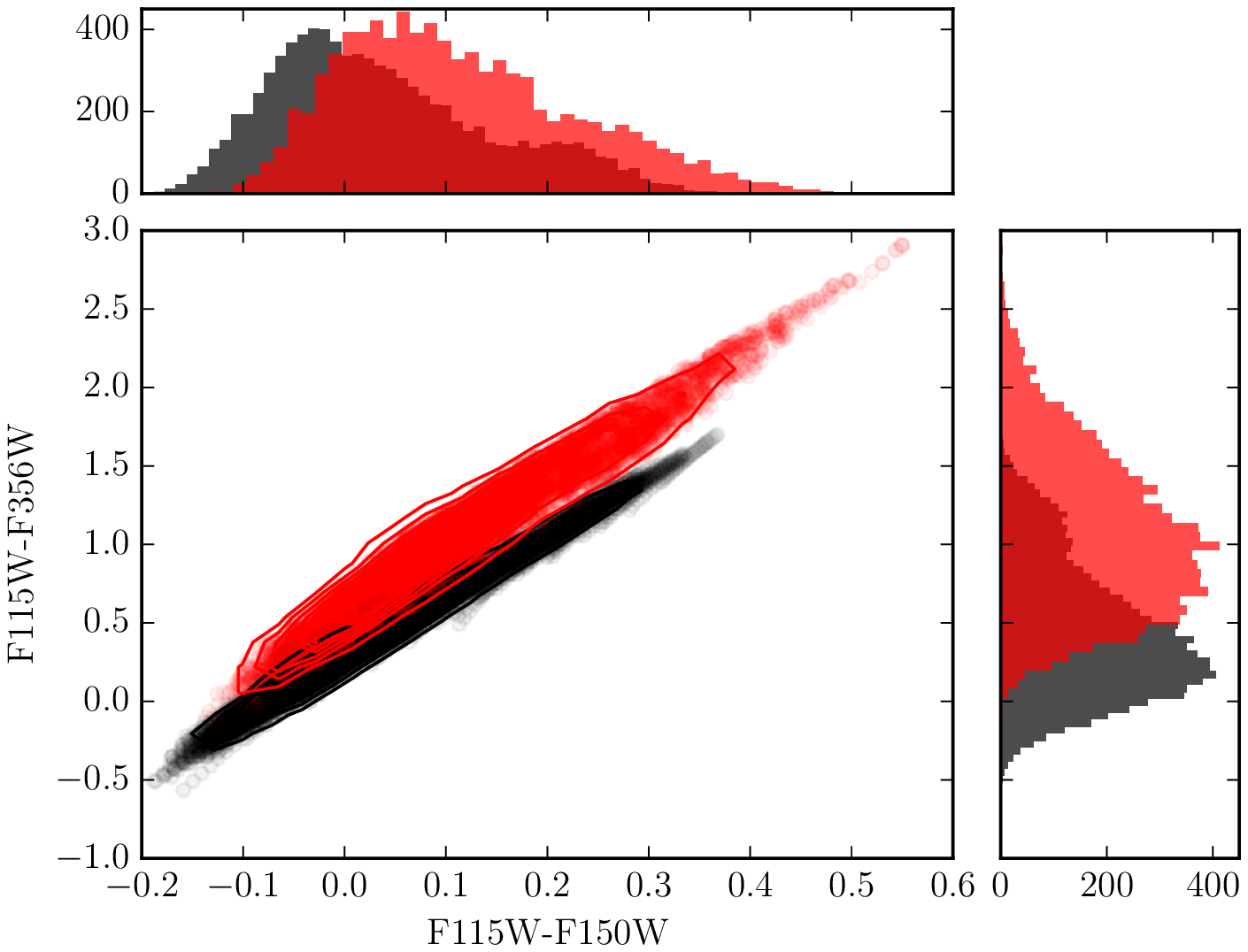}
		&
        \hspace{-18pt}\includegraphics[width=0.52\textwidth]{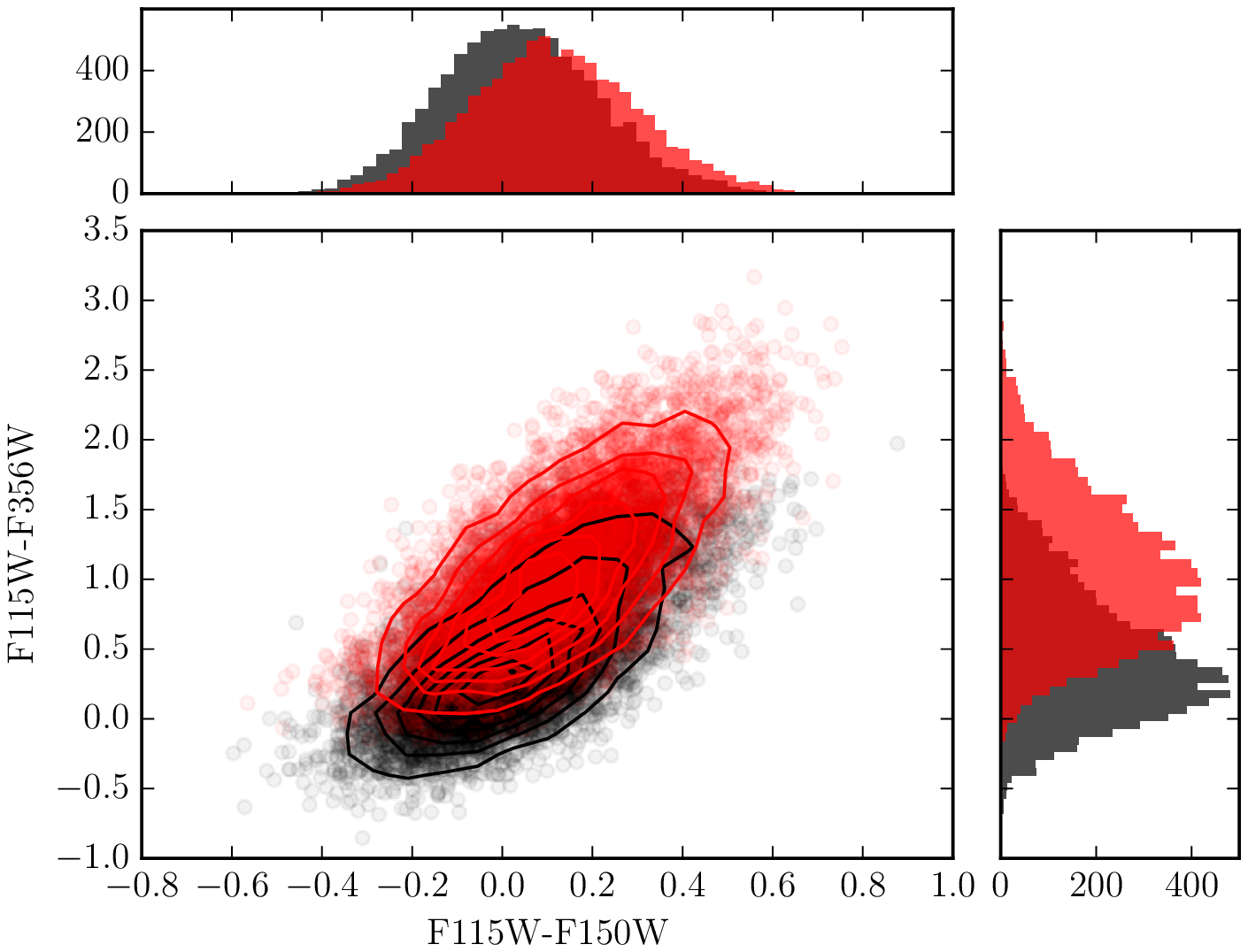}
	\end{tabular}
    \caption{Color-color diagrams showing the noise-free (left) distribution and the noisy ($\mathrm{SNR}=10$) distribution (right) of the quenched galaxies (red) and simulated (star-forming) galaxies (black).}
    \label{figure:ScatterHist}
\end{figure*}

In order to determine whether we can identify if a galaxy has undergone recent quenching or not, we use the calculated magnitudes in NIRCam and MIRI. The distribution of quenched and simulated galaxies in three filters is shown as a color-color diagram in \autoref{figure:ScatterHist}. In the noise-free case (panel 1, \autoref{figure:ScatterHist}) the quenched galaxies are largely separated from the normal galaxies. However, when adding noise in this sample, the overlap between the two classes becomes large (panel 2, \autoref{figure:ScatterHist}). In this case we assume that the extinction law is well characterized. We perform 10-fold cross validation with LDA on the whole set with all the available NIRCam filters and the MIRI F560W filter. This yields a mean accuracy or total accuracy of $\mathrm{TA} = 81 \pm 3 \%$ ($2\sigma$ error), where the accuracy for one validation iteration is defined as

\begin{figure}
	\centering 
        \includegraphics[width=\columnwidth]{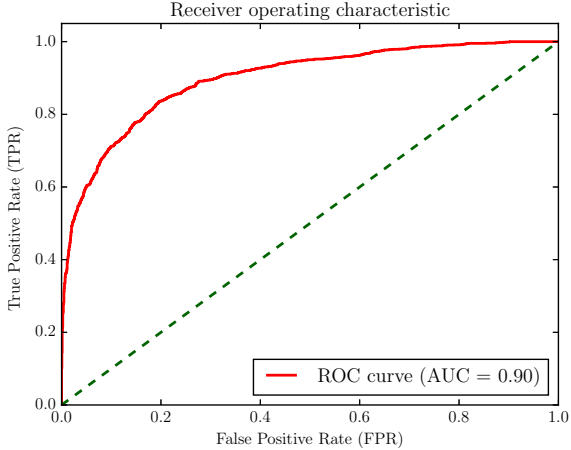}
    \caption{ROC curve for the fiducial set of galaxies. The red line gives the TPR versus the FPR (1-specificity) for the fiducial set as the threshold used for classification is changed. The dashed line displays the line of no-discrimination. The area under the ROC curve (AUC) is given in the lower right side of the figure.}
    \label{figure:ROC_curve}
\end{figure}

\begin{equation}
\label{eq:acc_eq}
\mathrm{A}=\frac{TQ+TN}{TQ+TN+FQ+FN}
\end{equation}
We define the recall, or true positive rate (TPR), the precision, or positive predictive value (PPV), where we use quenched as positive
\begin{equation}
\label{eq:TPR}
\mathrm{TPR}=\frac{TQ}{TQ+FN}
\end{equation}

\begin{equation}
\label{PPV}
\mathrm{PPV}=\frac{TQ}{TQ+FQ}
\end{equation}
And the corresponding quantities, specificity, or true negative rate (TNR), and negative predictive value (NPV) for the simulated star-forming, or `normal' galaxies

\begin{equation}
\label{eq:TNR}
\mathrm{TNR}=\frac{TN}{TN+FQ}
\end{equation}\textbf{}

\begin{equation}
\label{eq:NPV}
\mathrm{NPV}=\frac{TN}{TN+FN}
\end{equation}
where $TQ$ and $TN$ are true quenched and normal, respectively, (those galaxies that are correctly classified as either quenched or normal), and $FQ$ and $FN$ are false quenched and normal, respectively (i.e. those that are falsely identified as quenched and normal). \autoref{Tab:ConfusionMatrix_1} shows the confusion matrix when the model is applied to a test set. This is constructed by randomly taking out 20\% of the objects from the training data, meaning that the algorithm has never ``seen'' these objects before. In this case, 262 `normal' star-forming galaxies from the simulations are misclassified, while 1439 are correctly classified. The corresponding numbers for the quenched galaxies are 377 and 1333, which indicates that the algorithm performs slightly better on the normal galaxies with a specificity of approximately 85\% and a NPV of 79\%, while the recall for the quenched galaxies is 78\% and the precision is 84\%. This means that about 15\% of the normal galaxies will be incorrectly identified as quenched and about 22\% of the quenched galaxies will be incorrectly classified as normal. The receiver operating characteristic curve (ROC curve) in \autoref{figure:ROC_curve} displays how the TPR (recall) and the false positive rate (FPR) changes when changing the threshold used for the classification, together with the area under the curve (AUC). Ideally, one would want zero FPR and maximal TPR, which would mean you have a perfect classifier (i.e. you have no false positive and thus all your positives are true). The dashed line in the ROC curve (\autoref{figure:ROC_curve}) is the line of no discrimination, getting a ROC curve of a classifier along this line would mean our classifier is classifying at random. Having an AUC of unity corresponds to perfect classification, while 0.5 would correspond to a classifier which makes random guesses.

If testing is done only on galaxies that have $\EWHB<50$ \AA \ (which corresponds to galaxies with a 50\% escape fraction according to the method of \cite{zackrisson_spectral_2013,zackrisson_spectral_2017}) we get the confusion matrix shown in \autoref{Tab:ConfusionMatrix_2}. We see that the specificity (and NPV) in this case is approximately 87\% (82\%), while the recall (and precision) are 80\% (86\%).

The general trend of increasing the number of photometric filters used to train and test the algorithm is to increase the total accuracy. However, excluding the MIRI F560W filter from the classification does not significantly affect the result, indicating that it holds very little additional information about the quenching. 

\begin{figure}
	\centering 
        \includegraphics[width=\columnwidth]{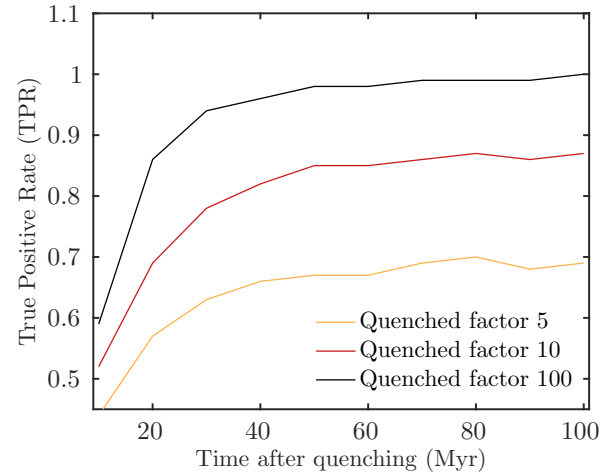}
    \caption{The recall (TPR) of the LDA model as a function of quenching duration for the different factors of quenching. Results shown here are for the fiducial set of objects where no limits have been put on the $\EWHB$.}
    \label{figure:LDA_Results}
\end{figure}

In order to identify the most problematic cases of quenching, and for identifying when the algorithm would break down, we create testing sets for each quenching factor and quenching duration. These sets are not used in the respective training phase, hence the algorithm has not `seen' the objects before. The algorithm is once again trained on a set of mixed quenched and simulated star-forming objects but now tested on a set consisting only of quenched galaxies with a certain quenching factor and quenching duration. The results of this test are shown in \autoref{figure:LDA_Results}. Not surprisingly, the algorithm performs better when quenching is stronger and the quenching time scale is longer, i.e. when the galaxies exhibit clearer signs of aging. If the galaxies are observed only 10 Myr after quenching takes place, the model only achieves a recall of 44-59\% depending on the level of quenching, meaning that these galaxies will generally not be distinguishable from simulated star-forming galaxies. Furthermore, it appears that galaxies that have undergone a weak quenching (factor 5 drop in SFR) are problematic.  However, for the most extreme cases of quenching (factor 100), a recall of 94\% is achieved even after a relatively short quenching duration ($\sim$ 30 Myr). The recall rate for galaxies that have undergone strong quenching (factor 10-100) after 50 Myr is $\gtrsim 85\%$.

\begin{table}
\centering
\caption{Confusion matrix giving the true class \emph{(true)} versus the class that the LDA algorithm predicts \emph{(predicted)} for a testing set consisting of 20\% of the total set.}
\label{Tab:ConfusionMatrix_1}
\begin{tabular}{c c c c }
\hline 
\hline
& & \multicolumn{2}{c}{Predicted} \\
& & Normal & Quenched \\
\hline
{\multirow{2}{*}{True}} & Normal & 1439 & 262\\
& Quenched & 377 & 1333 \\
\hline
\end{tabular}
\end{table}

\begin{table}
\centering
\caption{Confusion matrix showing the true class \emph{(true)} versus the class that the LDA algorithm predicts \emph{(predicted)} when the test set is limited to galaxies that have $\EWHB<50$ \AA}
\label{Tab:ConfusionMatrix_2}
\begin{tabular}{c c c c }
\hline 
\hline
& & \multicolumn{2}{c}{Predicted} \\
& & Normal & Quenched \\
\hline
{\multirow{2}{*}{True}} & Normal & 429 & 63\\
& Quenched & 97 & 395 \\
\hline
\end{tabular}
\end{table}

\subsection{Effects of dust handling}
\label{sec:dust_handling}
If we have access to JWST/NIRSpec observations, a correction for dust reddening could in principle be done for galaxies using $\mathrm{H}\beta$ and $\mathrm{H}\gamma$ for $z\sim 6-9$. However, in cases with high escape fractions ($\fesc \geq 0.5$), $\mathrm{H}\gamma$ is relatively weak ($\mathrm{EW(H\gamma)} < 30$ \AA) for a majority of the simulated galaxies. Furthermore, as mentioned in \autoref{sec:simulations}, the new version of BPASS binary evolution model predicts lower values of $\xiion$, which would lead to weaker emission lines. This and the fact that $\mathrm{H}\alpha$ will only be detectable out to $z\sim 7.2$ makes it unclear whether or not we can actually reliably correct for dust in these galaxies. We therefore assume that we cannot apply any dust correction.

In \autoref{Tab:Dustresults} we show results for the recall, precision, specificity and NPV when the algorithm is applied to sets with different attenuation/extinction laws. We see no significant impact on the performance when switching from testing and training on the Calzetti law where $\mathrm{E(B-V)_{stars}=E(B-V)_{neb}}$ to the standard Calzetti law where $\mathrm{E(B-V)_{stars}=0.44E(B-V)_{neb}}$. Furthermore, there is no significant difference in performance if the model is trained on either Calzetti law, and tested on the other. Meanwhile, the difference in performance between training on any of the Calzetti laws and testing on the steeper SMC law is larger, giving a slight increase in performance. We see an overall increase in performance if the model is trained and tested on the SMC law, implying that the steepness of the extinction law leads to less confusion between quenched and `normal' galaxies. However, we also see a significant decrease in the performance if one assumes a extinction law that is too steep, ie. if we train on the SMC law and apply the model to a population which is actually experiencing dust-reddening effects by a flatter Calzetti law. In this case, the algorithm will achieve a TNR of 66-67\%, which means that 33-34\% of `normal' star-forming galaxies will be wrongly identified as quenched. 

\begin{table}
	\centering
\caption{Recall (TPR), precision (PPV), specificity (TNR) and negative predictive value (NPV) for the LDA model when it is trained (T) and tested (P) on different dust laws. `Cal$^{\star}$' is the Calzetti law where $\mathrm{E(B-V)_{stars}=E(B-V)_{neb}}$, `Cal' is the standard Calzetti law and `SMC' is the \citet{pei_interstellar_1992} SMC extinction law}
\label{Tab:Dustresults}
\begin{tabular}{l c c c c}
\hline 
\hline
Set & TPR & PPV & TNR & NPV \\
 & (\%) & (\%) & (\%) & (\%) \\
\hline
T:Cal$^{\star}$ P: Cal$^{\star}$ & 78 & 84 & 85 & 79  \\
T:Cal$^{\star}$ P: Cal & 80 & 85 & 86 & 81  \\
T:Cal$^{\star}$ P: SMC & 84 & 87 & 88 & 85  \\
T:Cal P: Cal$^{\star}$ & 80 & 82 & 83 & 80  \\
T:Cal P: Cal & 80 & 86 & 87 & 81  \\
T:Cal P: SMC & 87 & 83 & 83 & 76  \\
T:SMC P: Cal$^{\star}$& 89 & 72 & 66 & 86  \\
T:SMC P: Cal & 90 & 73 & 67 & 87  \\
T:SMC P: SMC & 89 & 89 & 88 & 89  \\
\hline
\end{tabular}

\end{table}

\section{Discussion}
The previous sections provide evidence for the statement that rapid variations in the SFR in reionziation epoch galaxies will have an impact on the applicability of methods that utilize the emission lines and UV slope to estimate the escape fraction of LyC photons. Quenching of the type discussed in this paper will leave imprints on the SED that are in some degree degenerate with those produced by an extreme escape fraction of LyC photons. However, the effect that quenching has on the overall shape of the SED should in cases of large decreases in SFR (by a factor of 10-100) in the past 50-100 Myr, allow us to statistically identify quenched galaxies using JWST NIRCam and/or NIRSpec observations. For example, consider a situation where our algorithm correctly classifies galaxies in approximately 85\% of cases. If we then observe 100 galaxies and find that 15 of those are identified as quenched, it is hard to say anything about how common quenching is. However, if we observe 100 objects and find that 25-35 of those are identified as quenched, it is very likely that there exists a quenched sub-population. Furthermore, by considering a subsample of those galaxies that have H$\beta$ equivalent width ($\EWHB<50$ \AA) we should be able to statistically distinguish between extreme LyC leakage and a recent decrease in SFR. It is still unclear how common this type of quenching is in the early Universe, but if there exists a population of EoR galaxies with star formation histories similar to the the one recently claimed by \cite{hashimoto_onset_2018} for a $z\approx 9.1$ object, JWST should be able to identify this.

While a sub-population with strong variations in the SFH could be detectable using the JWST after roughly 50-100 Myr, distinguishing between cases where weak quenching has taken place (or cases where we catch a galaxy shortly after quenching) and a high escape fraction will be challenging. If we are able to identify a number of strongly quenched galaxies, it is however, not unlikely that there are galaxies that experience weaker quenching but that are not correctly identified. In principle, this should tell us how applicable the method presented in \cite{zackrisson_spectral_2013}, \cite{jensen_machine-learning_2016} and \cite{zackrisson_spectral_2017} is. If quenching is relatively common, it is likely that there will be cases in which the UV slope and emission lines will lead us to wrongly assume large leakage of LyC photons, and these cases may actually not be distinguishable using only JWST observations.

We have shown that the \cite{zackrisson_spectral_2013} method is subject to a degeneracy between the LyC escape fraction and star formation activity. Other methods to determine the escape fraction such as high $\left[ \Othree \right]/\left[ \Otwo \right]$ line ratios \cite{jaskot_origin_2013}, using UV absorption lines to probe the line-of-sight covering fraction \citep{jones_keck_2013,leethochawalit_absorption-line_2016} or constraining the escape fraction using Lyman-$\alpha$ profiles \citep{verhamme_using_2015} are not subject to the same degeneracy. These methods do, however, have other limitations that are important to consider when applying the methods to EoR galaxies, as discussed in \autoref{section:introduction}.

Meanwhile, identifying a quenched population of galaxies using JWST in itself also depends on our understanding of what dust extinction/attenuation law is most representative for the galaxy population at EoR. Even with the JWST/NIRSpec it not clear if we can reliably perform dust corrections for galaxies at $z\gtrsim 7$. A possible solution that has not been discussed here yet is the combination of MIRI and NIRSpec spectroscopic data. While NIRSpec will not be able to get spectra with H$\alpha$ at redshifts $z\gtrsim 7.2$, MIRI may be able to provide us with these data. Another possibility could be to characterize dust-reddening effects at lower redshifts, where NIRSpec will have coverage to simultaneously observe H$\alpha$, H$\beta$ and possibly H$\gamma$ and assume that this dust law is representative even at higher redshifts.
In short, any method that can characterize the dust-reddening effects and the variations in the reddening in galaxies during EoR will be extremely useful. If it is the case that there is no extinction law that is representative for the population of galaxies in the early universe, and in principle every galaxy has a different type of reddening at work, this will strongly limit the applicability of the method that we discuss here. Furthermore, determining which galaxies have undergone quenching also requires that we have a good understanding of how stars are formed and evolve from stellar evolutionary models.

We find that removing the MIRI F560W filter does not significantly affect the performance of the algorithm. The reason for this is most likely that the F560W filter does not probe a significantly different region than the NIRCam F444W filter, and thus provides little additional information about the overall shape of the SED. It is possible that the redder MIRI filters hold information that may improve the identification. However, the limited sensitivity in the F770W and redder filters makes it unclear if these will actually provide information that help the identification of quenched galaxies at these redshifts.

The classification algorithm can also produce results in terms of continuous class memberships (the likelihood for each single galaxy of belonging to a certain class). Computing this for an observed sample of galaxies, one could test how likely we are to find the observed distribution given only a distribution of non-quenched galaxies. This could be used to understand if a quenched sub-population exists in the observed sample of galaxies. However, performing formal significance tests (in the form of P-values) to these statements is challenging as it would involve estimating the underlying distribution of non-quenched galaxies.

We have also tried using a nonlinear support vector classifier \citep[][pp. 417-422]{hastie_elements_2009} with a radial basis function kernel and tuned parameters, and find no significant difference in performance compared to LDA.

\section{Summary \& Conclusions}
\begin{itemize}

\item We demonstrate that reionization-epoch galaxies that have experienced a recent decrease in star formation activity give rise to JWST spectra that are to some extent degenerate to those produced by extreme LyC leakage. This could seriously affect the applicability of the \cite{zackrisson_spectral_2013}, \cite{jensen_machine-learning_2016} and \cite{zackrisson_spectral_2017} method for identifying cases of high-LyC leakage based on low emission line equivalent widths with JWST/NIRSpec.

\item We show that if we have a good understanding of the general properties of the galaxy population during EoR, we should be able to statistically identify luminous galaxies that have undergone a significant decrease in SFR (by a factor of 10--100) in the past 50-100 Myr using JWST/NIRCam and/or NIRSpec observations. Cases with more moderate SFR fluctuations are more difficult to single out, but if there exists a significant population of $z>6$ galaxies with star formation histories similar to that recently claimed by \cite{hashimoto_onset_2018} for a $z\approx 9.1$ object, then JWST should easily be able to pick up on this. Conversely, it should be possible to -- in a statistical sense -- distinguish galaxies with extreme LyC leakage from objects that have experienced a significant drop in star formation activity.

\item While the UV/optical dust extinction is generally assumed to be quite low for $z>6$ galaxies, the slope of the dust extinction law represents one of the main uncertainties in attempts to break the degeneracy between extreme LyC leakage and quenched star formation. Hence, observations attempting to quantify the details of the dust-reddening effects at $z>6$ should be considered a high-priority task for JWST. 

\end{itemize}

\section*{Acknowledgements}
EZ acknowledges research grants from Swedish National Space Board and stiftelsen Olle Engkvist Byggm\"{a}stare.
IS is supported in part by the JSPS KAKENHI Grant Number JP26247022 and JP17H01111. Numerical simulations of IS were performed on Cray XC30 at at the Center for Computational Astrophysics, National Astronomical Observatory of Japan.




\bibliographystyle{mnras}
\bibliography{bibliography} 








\bsp	
\label{lastpage}
\end{document}